\documentclass[12pt,notoc]{JHEP3} 

\usepackage{epsfig,multicol,bbm,amsmath}

\newcommand\nn{\nonumber}
\newcommand{\de}{\delta}

\newcommand\fverb{\setbox\fverbbox=\hbox\bgroup\verb}
\newcommand\fverbdo{\egroup\medskip\noindent%
            \fbox{\unhbox\fverbbox}\ }
\newcommand\fverbit{\egroup\item[\fbox{\unhbox\fverbbox}]}
\newbox\fverbbox

\newcommand{\nablaslash}{\not{\hbox{\kern-3pt $\nabla$}}}

\title{Superconformal M2-branes and\\generalized Jordan triple systems}

\author{Bengt E.W.~Nilsson$\,{}^1$ and Jakob Palmkvist$\,{}^{12}$
\\\\
${}^1\,$Fundamental Physics\\
\phantom{${}^1\,$}Chalmers University of Technology\\
\phantom{${}^1\,$}SE-412 96 G\"oteborg, Sweden\\\\
${}^2\,$Max Planck Institute for Gravitational Physics\\
\phantom{${}^2\,$}Albert Einstein Institute\\
\phantom{${}^2\,$}Am M\"uhlenberg 1\\
\phantom{${}^2\,$}DE-14476 Golm, Germany\\\\
{\tt {\footnotesize tfebn@chalmers.se, jakob.palmkvist@aei.mpg.de}}}

\preprint{
AEI-2008-055
}

\abstract{Three-dimensional conformal theories with six
supersymmetries and $SU(4)$  $R$-symmetry describing stacks of
M2-branes are here proposed to be related to generalized Jordan
triple systems. Writing the four-index structure constants in an
appropriate form, the Chern-Simons part of the action immediately
suggests a connection to such triple systems. In contrast to the
previously considered three-algebras, the additional structure of a
generalized Jordan triple system is associated to a graded Lie
algebra, which corresponds to an 
extension of the gauge group. In this note we
show that the whole theory with six manifest supersymmetries can be
naturally expressed in terms of such a graded Lie algebra. Also the
BLG theory with eight supersymmetries is included as a special
case.}


\begin{document}


\setcounter{page}{2}

\section{Introduction}

A three-dimensional maximally ($\mathcal N=8$) superconformal theory
was recently constructed by Bagger, Lambert and Gustavsson (BLG) in
\cite{Bagger:2006sk,Gustavsson:2007vu,Bagger:2007jr,Bagger:2007vi}.
 The BLG theory was originally proposed to describe multiple M2-branes.
An interesting aspect of this theory is that it contains a
Chern-Simons term \cite{Schwarz:2004yj} making the BLG theory
potentially interesting also for  condensed matter applications. The
multiple M2-brane interpretation has, however, met with a number of
problems having to do with the algebraic structure on which the
theory is based. The theory contains a kind of four-index structure
constant for a three-algebra with a Euclidean metric. This
three-algebra has, however, been proven
\cite{Papadopoulos:2008sk,Gauntlett:2008uf} to have basically only
one realization, $\mathcal A_4$, related to the ordinary Lie algebra
$so(4)$ through its totally antisymmetric epsilon tensor. This is
limiting the role of the BLG theory to stacks of two M2-branes
\cite{Lambert:2008et,Distler:2008mk}.

By relaxing the assumption that the metric on the algebra should be
positive definite \cite{Gran:2008vi} any Lie algebra can be
accommodated. The drawback of using a degenerate metric as done in
\cite{Gran:2008vi} is that it produces a set of field equations
which cannot be integrated to a Lagrangian if the zero norm mode is
not assumed constant. This subsequently led to a number of attempts
to use a non-degenerate but Lorentzian metric
\cite{Gomis:2008uv,Benvenuti:2008bt,Ho:2008ei}. Again there are
problems; these theories make sense only provided the negative norm
modes can be rendered harmless. Even when this is the case they are
of real interest only if they contain genuine M2-physics instead of
just providing a reformulation of the D2-brane. For some recent
results in this direction, see
\cite{Bandres:2008kj,Gomis:2008be,Ezhuthachan:2008ch,Cecotti:2008qs}.

From the work of  \cite{Gran:2008vi} it was also clear that the
structure constants need not be totally antisymmetric. This might be
interesting since this property seems to be part of the reason why
only one realization, related to $SO(4)$,  of the fundamental
identity can be constructed in the Euclidean case. In fact, as
realized by Aharony, Bergman, Jafferis and Maldacena (ABJM)
\cite{Aharony:2008ug}, by reducing the number of linearly realized
supersymmetries from the maximal $\mathcal N=8$ to $\mathcal N=6$
this no-go theorem can be avoided. Following
\cite{VanRaamsdonk:2008ft}, the authors of \cite{Aharony:2008ug}
(see also \cite{Benna:2008zy,Bandres:2008ry}) used a construction
with the fields in the bi-fundamental representation of $U(N) \times
U(N)$ and without any reference to the four-index structure
constants. However, in a work following this Bagger and Lambert
\cite{Bagger:2008se} pointed out that if reinstating the four-index
structure constants there are interesting implications for their
antisymmetry properties. In particular, six supersymmetries are
compatible with structure constants which are not totally
antisymmetric.

The purpose of this note is to write the  structure constants in yet
another form which suggests the possibility of relating them to
certain algebraic structures, known as
generalized Jordan triple systems.
Since this result will rely on embeddings into infinite dimensional graded Lie algebras $g$ we should here mention that embeddings into finite dimensional ones are also possible \cite{Gaiotto:2008sd,Hosomichi:2008jd,Hosomichi:2008jb,Schnabl:2008wj} but then $g$ is a (three graded) Lie superalgebra.

The paper is organized as follows. In section two we review the ABJM
theory and present the Lagrangian in terms of four-index structure
constants as described in \cite{Bagger:2008se}.  In section three we
then provide a reformulation of this theory in terms of structure constants adapted to
triple systems. Some relevant aspects of generalized Jordan triple
systems
and the associated graded Lie algebra
are summarized in section four. The
last section contains conclusions and some further comments.

\section{The ABJM M2-theory}

The BLG theory contains three different fields; the two propagating
ones $X^{I}{}_{a}$ and $\Psi_{a}$, which are three-dimensional scalars
and spinors, respectively, and the auxiliary gauge field
$\tilde{A}_{\mu}{}^{a}{}_{b}$. Here the indices $a,\,b,\,\ldots$ are connected to
the three-algebra and some $n$-dimensional basis $T^{a}$, while the
$I,\,J,\,K,\,\ldots$ indices are $SO(8)$ vector indices. The spinors
transform under a spinor representation of  $SO(8)$ but the
corresponding index is not written out explicitly. Indices $\mu,\,\nu,\,
\ldots$ are vector indices on the flat M2-brane world volume.

Using these fields one can write down $\mathcal N=8$ supersymmetry
transformation rules and  covariant field equations.  This is
possible without introducing a metric on the three-algebra. In such
a situation the position of the indices on the structure constants
is fixed as $f^{abc}{}_d$. The corresponding fundamental identity
needed for supersymmetry and gauge invariance then reads
\cite{Bagger:2006sk,Gustavsson:2007vu,Bagger:2007jr,Bagger:2007vi},
\begin{equation}
f^{abc}{}_g f^{efg}{}_d = 3f^{ef[a}{}_g f^{bc]g}{}_d  \,, \label{FI}
\end{equation}
which can be written in the following alternative but equivalent
form \cite{Gran:2008vi},
\begin{equation}
f^{[abc}{}_g f^{e]fg}{}_d = 0 \,. \label{WFI}
\end{equation}

The construction of a Lagrangian requires the introduction of a
metric on the three-algebra. As discussed above,  if one wants to
describe more general Lie algebras than $so(4)$,  this metric must
be degenerate \cite{Gran:2008vi} or non-degenerate but indefinite
\cite{Gomis:2008uv,Benvenuti:2008bt,Ho:2008ei}. Finally, to
construct an action one also needs to introduce the basic gauge
field $A_{\mu ab}$ \footnote{However, already gauge invariance of
the field equations requires this gauge field \cite{Gran:2008vi}.}
which is related to the previously defined gauge field and structure
constants as follows:
\begin{equation}
\tilde{A}_{\mu}{}^{a}{}_b=A_{\mu cd}f^{cda}{}_{b}\,.
\end{equation}

The BLG Lagrangian is
 \cite{Bagger:2007jr}
\begin{eqnarray}
{\cal L} &=& -\tfrac{1}{2}(D_\mu X^{Ia})(D^\mu X^I{}_a) + \tfrac{i}{2}
\bar \Psi^a \gamma^\mu D_\mu \Psi_a +
 \tfrac{i}{4} \bar\Psi_b \Gamma_{IJ} X^I{}_c X^J{}_d \Psi_a f^{abcd}
\nonumber\\
&&-V +\tfrac{1}{2}\varepsilon^{\mu\nu\lambda}\left( f^{abcd}A_{\mu ab}\partial_\nu A_{\lambda cd} +
 \tfrac{2}{3} f^{cda}{}_g f^{efgb} A_{\mu ab}  A_{\nu cd} A_{\lambda ef}  \right) \,,
\end{eqnarray}
where the potential is given by
\begin{equation}
V = \tfrac{1}{12} f^{abcd}f^{efg}{}_d X^I{}_a X^J{}_b X^K{}_c X^I{}_e X^J{}_f X^K{}_g\,.
\end{equation}
Note that in terms of $\tilde A$ the Chern-Simons term becomes
\begin{equation}
{\cal L}_{CS}=\tfrac{1}{2}\varepsilon^{\mu\nu\lambda}\left( A_{\mu
ab}\partial_\nu \tilde A_{\lambda}{}^{ab} + \tfrac{2}{3} A_{\mu}{}^a{}_b
\tilde A_{\nu}{}^b{}_c \tilde A_{\lambda}{}^c{}_{a}  \right)
\end{equation}
and that the fundamental identity implies that, in the variation of the last term,
the structure constants can be associated with any two of the three
vector fields.

Following ABJM \cite{Aharony:2008ug} we now rewrite this in a form
which has only six manifest supersymmetries and manifest $SU(4)$
$R$-symmetry. As emphasized by these authors, this is naturally done
using matter fields in the bi-fundamental representation
\cite{VanRaamsdonk:2008ft} of $U(N)\times U(N)$, and no reference to
three-algebras and their structure constants is needed. However, for
the purpose of this note we need to reinstate the four-index
structure constants. Fortunately, this was discussed in detail in a
recent work by Bagger and Lambert \cite{Bagger:2008se}.

The ABJM action is expressed in terms of complex scalar fields
$Z^A{}_a$ and spinors $\Psi_{Aa}$ with the capital indices
transforming in fundamental and anti-fundamental representations of
the $SU(4)$ $R$-symmetry, respectively.
If rewritten in terms of
four-index structure constants as done in \cite{Bagger:2008se} 
(but rescaled by a factor of two), the
ABJM action reads
\begin{eqnarray}
{\cal L} &=& -(D_\mu Z^{A}{}_a)(D^\mu \bar{Z}_A{}^a) - i \bar
\Psi^A{}_a \Gamma^\mu
 D_\mu  \Psi_A{}^a
 \nonumber\\
 &&  - i f^{abcd}\bar {\Psi}^A{}_d  \Psi_{Aa}Z^B{}_b \bar{Z}_{Bc}+
 2i f^{abcd}\bar {\Psi}^A{}_d  \Psi_{Ba}Z^B{}_b \bar{Z}_{Ac}
 \nonumber \\
 &&
 - \tfrac{i}{2}\epsilon_{ABCD} f^{abcd}\bar {\Psi}^A{}_c  \Psi^B{}_d Z^C{}_a Z^D{}_b
  -\tfrac{i}{2}\epsilon^{ABCD} f^{cdab}\bar {\Psi}_{Ac}  \Psi_{Bd}\bar{Z}_{Ca}\bar{Z}_{Bd}
\nonumber\\
&&
-V +\tfrac{1}{2}\epsilon^{\mu\nu\lambda}( f^{abcd}A_{\mu ab}\partial_\nu A_{\lambda cd} +
\tfrac{2}{3} f^{cda}{}_g f^{efgb} A_{\mu ab}  A_{\nu cd} A_{\lambda ef} ) \,,
\end{eqnarray}
where the potential can be written
\begin{equation}
V = \tfrac{2}{3} \Upsilon^{CD}{}_{Bd}\bar\Upsilon_{CD}{}^{Bd} \,,
\end{equation}
\begin{equation}
  \Upsilon^{CD}{}_{Bd}= f^{abc}{}_d Z^C{}_a{Z}^D{}_b \bar{Z}_{Bc}
  + f^{abc}{}_d\delta^{[C}{}_B Z^{D]}{}_a{Z}^E{}_b \bar{Z}_{Ec} \,.
\end{equation}
In order to write this action one needs a metric on the
three-algebra to raise and lower three-algebra indices. The structure
constants appearing in this formulation of the $\mathcal N=6$ ABJM
theory \cite{Bagger:2008se} are antisymmetric in the first pair of
indices as well as in the second pair while complex conjugation is
defined to interchange the two pairs of indices.

As we will see below the need for an explicit metric in the Lagrangian can be eliminated by
writing the structure constants as $f^{ab}{}_{cd}$ or
$f^{a}{}_{b}{}^{c}{}_{d}$ (which we will see later are in fact
related to each other). This will also require the introduction of a graded Lie algebra in
a way that will be explained in the next section.

\section{Structure constants adapted to triple systems}

Our next goal is to try to relate the M2-brane to generalized
Jordan triple systems. The first step is to rewrite the $\mathcal N$= 6
M2-theory as formulated at the end of the previous section in terms of
structure constants with two upper
and two lower indices, which are antisymmetric in each pair separately,
\begin{equation}
f^{ab}{}_{cd}=f^{[ab]}{}_{cd}=f^{ab}{}_{[cd]}\,.
\end{equation}

The crucial difference between our approach and the one used in
\cite{Bagger:2008se} is that we do not consider the fields $Z^A,\,\Psi_A$
as elements in the same three-algebra as their complex conjugates
$\bar{Z}_A,\,\Psi^A$. (We save the bar on the spinor for the Dirac conjugate.) Rather, we are dealing with two vector spaces $g_1$ and $g_{-1}$, with bases $T^a$ and $T_a$, respectively. These two vector spaces generate a graded Lie algebra $g$. We do not use any metric on $g_1$ and $g_{-1}$ to raise and lower indices, but we use an antilinear involution $\tau$ on $g$ to go between the subspaces,
$\tau(T^a)=T_a$. We also use a bilinear form
on $g$ to contract upper and lower indices.
We will describe this graded Lie algebra in more detail in the next section. Here we just
define the components of the fields $Z^A,\,\Psi_A$ in $g_1$ to have the index structure
$Z^A{}_a,\,\Psi_{Aa}$. 
The components of $\tau(Z^A),\,\tau(\Psi_A)$ in $g_{-1}$ are then
the complex conjugates
$\bar Z_A{}^a,
 \Psi^{Aa}$. 
That it is
natural to place the indices like this can be seen from rewriting
the Bagger-Lambert version of the ABJM action as follows:
\begin{eqnarray}
{\cal L} &=& -(D_\mu Z^{A}{}_a)(D^\mu \bar{Z}_A{}^a) - i \bar
\Psi^{Aa} \gamma^\mu
 D_\mu  \Psi_{Aa}
 \nonumber\\
 &&  - i f^{ab}{}_{cd}\bar {\Psi}^{Ad}
 \Psi_{Aa}Z^B{}_b \bar{Z}_{B}{}^{c}+
 2i f^{ab}{}_{cd}\bar {\Psi}^{Ad}  \Psi_{Ba}Z^B{}_b \bar{Z}_{A}{}^{c}
  \nonumber\\
 &&
 -\tfrac{i}{2}\epsilon_{ABCD} f^{ab}{}_{cd}\bar {\Psi}^{Ac}  \Psi^{Bd} Z^C{}_a Z^D{}_b
  -\tfrac{i}{2}\epsilon^{ABCD} f^{cd}{}_{ab}\bar {\Psi}_{Ac}  \Psi_{Bd}\bar{Z}_{C}{}^a \bar{Z}_{D}{}^{b}
\nonumber\\
\label{lagrange} && -V +\tfrac{1}{2}\epsilon^{\mu\nu\lambda}(
f^{ab}{}_{cd}A_{\mu}{}^d{}_{b}
\partial_\nu A_{\lambda}{}^c{}_{a}+ \tfrac{2}{3} f^{bd}{}_{gc} f^{gf}{}_{ae}
A_{\mu}{}^a{}_{b}  A_{\nu}{}^c{}_{d} A_{\lambda}{}^e{}_{f}) \,,
\end{eqnarray}
where the potential now takes the form
\begin{equation}
V = \tfrac{2}{3} \Upsilon^{CD}{}_{Bd}\bar\Upsilon_{CD}{}^{Bd} \,,
\end{equation}
\begin{equation}
  \Upsilon^{CD}{}_{Bd}= f^{ab}{}_{cd} Z^C{}_a{Z}^D{}_b \bar{Z}_B{}^c
  + f^{ab}{}_{cd}\delta^{[C}{}_B Z^{D]}{}_a{Z}^E{}_b \bar{Z}_{E}{}^{c} \,.
\end{equation}
This action can be shown to be $\mathcal N=6$ supersymmetric provided
that
the
structure constants obey
\begin{equation} \label{fundid}
f^{a[b}{}_{dc} f^{e]d}{}_{gh} = f^{be}{}_{d[g} f^{ad}{}_{h]c}\,
\end{equation}
and, under complex conjugation,
\begin{equation} \label{komplexkonjugering}
(f^{ab}{}_{cd})^\ast = f^{cd}{}_{ab} \equiv f_{ab}{}^{cd}
\end{equation}
One immediate way to see that this identity is relevant is to consider the Chern-Simons term
\begin{equation}
{\cal L}_{CS}=\tfrac{1}{2}\varepsilon^{\mu\nu\lambda}\left( A_{\mu}{}^{b}{}_a\partial_\nu \tilde A_{\lambda}{}^a{}_b + \tfrac{2}{3} A_{\mu}{}^a{}_{b} \tilde A_{\nu}{}^b{}_{c} \tilde A_{\lambda}{}^c{}_a  \right),
\end{equation}
where we use vector fields $A_{\mu}{}^a{}_{b}$ and
\begin{equation} \label{treelva}
\tilde A_{\mu}{}^a{}_{b} =f^{ac}{}_{bd} A_{\mu}{}^d{}_{c}\,.
\end{equation}
The identity (\ref{fundid}) then follows from the
observation that when deriving the field equation the variation of
each vector field must provide an identical contribution to the
answer. Note that also the Chern-Simons field without tilde has an
upper and a lower index which is not the case in previous treatments
of the M2-brane system.

As we will see in the next section, the structure constants can
also be written as $f_a{}^b{}_c{}^d$. It is then interesting to note
that they, as well as their corresponding fundamental identity,
appear naturally also in the embedding tensor formalism of
\cite{Bergshoeff:2008bh} but for seemingly completely different
reasons.

In terms of structure constants of generalized Jordan triple systems the transformation
rules for the six supersymmetries, parametrized by the complex
self-dual three-dimensional spinor $\epsilon_{AB}$, read
\begin{equation}
\delta Z^A{}_a=i \bar \epsilon^{AB}\Psi_{Ba}\,,
\end{equation}
\begin{equation}
\delta \Psi_{Bd}=\gamma^\mu D_\mu Z^A{}_d \epsilon_{AB} + f^{ab}{}_{cd} Z^C{}_a Z^D{}_b \bar Z_B{}^c
 \epsilon_{CD}- f^{ab}{}_{cd} Z^A{}_a Z^C{}_b \bar Z_C{}^c \epsilon_{AB}\,,
\end{equation}
while the Chern-Simons one-form transforms as follows:
 \begin{equation}
\delta A_\mu{}^a{}_{b}=-i \bar \epsilon_{AB} \gamma_\mu \Psi^{Aa} Z^B{}_b
+
 i \bar \epsilon^{AB} \gamma_\mu \Psi_{Ab} \bar Z_{B}{}^{a}\,.
\end{equation}
To prove that the Lagrangian has six supersymmetries only requires
the use of the identities (\ref{fundid}) and (\ref{komplexkonjugering}).
The latter is needed since $\tau$ is antilinear. We have for example
\begin{align}
\tau(f^{ab}{}_{cd}Z^A{}_e)=(f^{ab}{}_{cd})^\ast \tau(Z^A{}_e)
=f^{cd}{}_{ab} \bar Z_A{}^e.
\end{align}
This also ensures that the kinetic term in the Lagrangian is positive-definite.
In order to see how the identity (\ref{fundid}) arises in generalized Jordan triple systems, we need to discuss some
further aspects of the underlying graded Lie algebra.

\section{Triple systems and graded Lie algebras}
\label{trippgrad}
In this section we will describe how the two vector spaces $g_1$ and $g_{-1}$, with bases $T^a$ and $T_a$, respectively, generate a graded Lie algebra $g$. The fact that $g$ is graded means that $g$ can be written as a direct sum of   
subspaces $g_k$ for all integers $k$, such that
\begin{align} \label{gradering}
[g_i,\,g_j]\subseteq g_{i+j}
\end{align}
for all integers $i,\,j$ (with the possibility that $g_k = 0$ for all sufficiently large $|k|$).
We call $k$ the \textit{level} of the elements in $g_k$.

It follows in particular that any subspace $g_k$ form a representation of the subalgebra $g_0$.  
First we consider as $g_0$ the Lie algebra $sl(n)$, with generators $K^a{}_b$
and commutation relations
\begin{align}
[K^a{}_b,\,K^c{}_d]=\de^c{}_b K^a{}_d - \de^a{}_d K^c{}_b.
\end{align}
We let $sl(n)$ act on $g_1$ and $g_{-1}$ in the fundamental and antifundamental representation, respectively:
\begin{align} \label{kommrel1}
[K^a{}_b,\,T^c]&=\de_b{}^c T^a, & [K^a{}_b,\,T_c]&=-\de^a{}_c T_b\,.
\end{align}

In the graded Lie algebra we must also have 
$[g_{-1},\,g_1] \subseteq g_0$. For this we introduce the structure constants $f^a{}_b{}^c{}_d$ by
\begin{align} \label{kommrel2}
[T^a,\,T_b]&=f^a{}_b{}^c{}_d K^d{}_c \equiv S^a{}_b,
\end{align}
and from (\ref{kommrel1}) we get 
\begin{align} \label{kommrel3}
[S^a{}_b,\,T^c]&=f^a{}_b{}^c{}_d T^d, & [S^a{}_b,\,T_c]&=
-f^a{}_b{}^d{}_c T_d\,.
\end{align}
We thus have $[[T^a,\,T_b],\,T^c]=f^a{}_b{}^c{}_d T^d$, and analogously we define the structure constants $f_a{}^b{}_c{}^d$ by  
$[[T_a,\,T^b],\,T_c]=f_a{}^b{}_c{}^d T_d$. It follows from
(\ref{kommrel3}) that 
\begin{align}
f_a{}^b{}_c{}^d=f^b{}_a{}^d{}_c\,.
\end{align}
For the Jacobi identity
\begin{align}
[[T^a,\,T_b],\,T^c]-[[T^c,\,T_b],\,T^a]
=[[T^a,\,T^c],\,T_b]
\end{align}
to hold,
the structure constants must satisfy the identity
\begin{align}
f^a{}_b{}^f{}_g f^c{}_d{}^e{}_f=
f^c{}_d{}^f{}_g f^a{}_b{}^e{}_f+
f^a{}_b{}^c{}_f f^f{}_d{}^e{}_g-
f_b{}^a{}_d{}^f f^c{}_f{}^e{}_g\,.\label{generalized Jordan triple systemid}
\end{align}

We can now redefine $g_0$ to be the 
subalgebra of $sl(n)$ spanned by all elements $S^a{}_b$ 
(so that $[g_1,\,g_{-1}]=g_0$)
with the commutation relations
\begin{align}
[S^a{}_b,\,S^c{}_d]
=f^{a}{}_b{}^c{}_e S^e{}_d
-f_b{}^a{}_d{}^e S^c{}_e
\,.
\end{align}

Let $\tau$ be the restriction of the Chevalley involution on $sl(n)$ to $g_0$. Then $\tau$ can be extended by $\tau(T^a)=T_a$ to a graded involution on the whole of $g$, such that $\tau(g_k) = g_{-k}$ for all integers 
$k$. It follows from this property, together with (\ref{gradering}), that
$g_1$ closes under the triple product
\begin{eqnarray}
(abc)=[[a,\tau{(b)}],c],
\end{eqnarray}
(and likewise for $g_{-1}$). Thus 
the identity (\ref{generalized Jordan triple systemid}) can be expressed as
\begin{eqnarray} \label{jordit}
(ab(xyz))-(xy(abz))=((abx)yz)-(x(bay)z)\,.
\end{eqnarray}
This is the definition of a \textit{generalized Jordan triple
system}, a vector space with a triple product that satisfies (\ref{jordit}). Thus any graded Lie algebra with a graded involution
leads to a generalized Jordan triple
system. 
Conversely, for any
generalized Jordan triple system $T$,
there is an associated graded Lie algebra $g$, which is an extension of the vector space $g_{-1}+g_0+g_1$ that we described above \cite{Kantor,Kantor3.5,Asano2,Palmkvist:2007as}. 

We stress that
the Lie algebra associated to a generalized Jordan triple system is the whole graded Lie algebra $g$, and not only the subalgebra $g_0$, which (in the case of three-algebras) was
called `the associated Lie algebra' by
Bagger and Lambert in \cite{Bagger:2007jr}. 
The graded Lie algebra $g$ associated to a generalized Jordan triple system $T$ was constructed by Kantor in a way such that if $g$ is finite-dimensional, then
simplicity of $g$ is equivalent to \textit{K-simplicity} of $T$ \cite{Kantor,Kantor3.5}. A generalized Jordan triple system $T$ is K-simple if there is no proper non-trivial subspace $U$
such that $(TTU) \subseteq U$ and $(UTT) \subseteq U$. 

In the construction of the graded Lie algebra associated to a generalized Jordan triple system,
one defines generators
$T^{ab}=[T^a,\,T^b]$ at level two, $T^{abc}=[[T^a,\,T^b],\,T^c]$ at level
three, and so on, (and likewise $T_{ab},\,T_{abc},\,\ldots$ at the negative levels).
These elements will satisfy
\begin{align}
0=T^{(ab)cd\cdots}=T^{[abc]d\cdots}\,,
\end{align}
due to antisymmetry of the Lie bracket and the Jacobi
identity, but also further conditions 
that amount to factoring out
ideals from the free Lie algebra
generated by $T^a$ and $T_a$. 

Assume that $g_0$ is semisimple. We can extend the Killing form $\kappa$ on $g_0$ to the vector space $g_{-1}+g_0+g_1$ by $\kappa(T^a,\,T_b) = \delta^a{}_b$.
Then we can recursively (using the invariance) extend it to an invariant bilinear form on the whole of the free Lie algebra generated by $T^a$ and $T_a$,
provided that the structure constants satisfy
\begin{align} \label{kappavillkor}
f^a{}_b{}^c{}_d=f^c{}_d{}^a{}_b\,.
\end{align}
But this invariant bilinear form will then be degenerate, and the corresponding ideals are exactly the ones that we have to factor out to obtain the Lie algebra associated to the generalized Jordan triple system.
This can be done recursively.
Suppose that the restriction of the bilinear form to
the vector space $g_{-k+1} + \cdots +g_{k-1}$, for some $k$, is non-degenerate.
Then
$\kappa( x ,\, y )$, where $x \in g_{k}$ and $y \in g_{-k}$, is a linear
combination of terms
\begin{align}
f^{a_1\cdots  a_{k}}{}_{b_1\cdots b_{k}} \equiv
(-1)^{k+1}\kappa( T^{a_1\cdots  a_{k}} ,\,  T_{b_k\cdots b_{1}} ) = 
(-1)^{k}\kappa( [T^{a_1\cdots a_k},\,T_{b_{1}}] ,\,  T_{b_k \cdots
b_{2}} )\,.
\end{align}
Using the structure constants for the triple product, this can be
evaluated as
\begin{align} \label{massaindex}
f^{a_1\cdots a_k}{}_{b_1\cdots b_{k}} =f^{a_1}{}_{b_1}{}^{a_2}{}_c f^{c a_3\cdots a_k}{}_{b_2
\cdots b_k}-\sum f^{a_j}{}_{b_1}{}^{a_i}{}_c f^{c_{ij}}{}_{ b_2 b_3
\cdots b_k}\,,
\end{align}
where the sum goes over all $i,\,j$ such that $1 \leq i < j \leq k$
and $c_{ij}$ denotes the sequence of indices obtained from $a_1
\cdots a_{n}$ by omitting $a_j$ and replacing $a_i$ by $c$, that is,
\begin{align}
c_{ij}=a_1 \cdots a_{i-1}\,c\,a_{i+1} \cdots a_{j-1}\,a_{j+1} \cdots
a_{k}\,.
\end{align}

From now on, we assume that the structure constants are antisymmetric
in the first and third index. It then follows from (\ref{kappavillkor})
that they are antisymmetric also in the second and the fourth index:
\begin{align}
f^a{}_b{}^c{}_d = - f^c{}_b{}^a{}_d = - f^a{}_d{}^c{}_b\,
\end{align}
and we have
\begin{align} \label{frelation}
f^{ab}{}_{cd} = 2f^a{}_c{}^b{}_d
\end{align}
from the Jacobi identity.
The
identity (\ref{generalized Jordan triple systemid}) then becomes
\begin{equation}
 f^{e[a}{}_{dc}  f^{b]d}{}_{gh}=  f^{ab}{}_{d[g}  f^{ed}{}_{h]c}\,,
\end{equation}
which is precisely the identity needed in the previous section to
prove supersymmetry. Furthermore, we have
\begin{align}
(f^a{}_b{}^c{}_d)^\ast &= \kappa(\tau(f^a{}_b{}^c{}_e T^e),\,T^d)
=\kappa(\tau([[T^a,\,T_b],\,T^c]),\,T^d)\nn\\
&=\kappa([[T_a,\,T^b],\,T_c],\,T^d)=f_a{}^b{}_c{}^d
=f^b{}_a{}^d{}_c
\end{align}
since $\tau$ is antilinear, and using (\ref{frelation}) we get
$(f^{ab}{}_{cd})^\ast = f^{cd}{}_{ab}$.
Thus the requirements for the six
supersymmetries of the action (\ref{lagrange}) are satisfied.

With the antisymmetry $f^a{}_b{}^c{}_d=-f^c{}_b{}^a{}_d$, the first
term on the right hand side of (\ref{massaindex}) coincide with the first term in the summation.
In the case $k=3$ the equation simplifies to
\begin{align} \label{sexindex}
f^{abc}{}_{def}&=2f^{a}{}_{d}{}^b{}_{g}f^{gc}{}_{ef}
-f^{c}{}_{d}{}^a{}_{g}f^{gb}{}_{ef}-f^{c}{}_{d}{}^b{}_{g}f^{ag}{}_{ef}\nn\\
&=f^{ab}{}_{dg}f^{gc}{}_{ef}-\tfrac12f^{ca}{}_{dg}f^{gb}{}_{ef}
-\tfrac12f^{cb}{}_{dg}f^{ag}{}_{ef}
\nn\\
&= f^{ab}{}_{dg}f^{gc}{}_{ef}-f^{c[a}{}_{gd}f^{b]g}{}_{ef}\,.
\end{align}
We see that $f^{abc}{}_{def}$ is
antisymmetric in the first two indices and vanishes upon
antisymmetrization in the three upper indices (or the three lower
ones). This is in accordance with the Jacobi identity, since by definition
\begin{align}
f^{abc}{}_{def}=\kappa([[T^a,\,T^b],\,T^c],\,[[T_f,\,T_e],\,T_d])\,.
\end{align}
Continuing in this way, one can determine which symmetries the tensors
at each level must have, and their commutation relations follow
from the Jacobi identity. Thus the graded Lie algebra $g$ is completely determined by the generalized Jordan triple system $T$, or equivalently, by the structure constants $f^a{}_b{}^c{}_d$. 

We will now discuss some further properties of the Lie algebra $g$. We will make use of the following two theorems by Kantor, the first of which we mentioned already in the beginning of this section.
\newtheorem{thm}{Theorem} 
\begin{thm} {\rm\cite{Kantor3.5} (Section 3, Prop.~7$'$ and Theorem 1$'$. See also \cite{Asano2} Theorem 3.5.)}
\label{sats1}Assume that $g$ is finite-dimensional. Then $g$ is simple if and only if $T$ is $K$-simple. 
\end{thm}
\begin{thm} {\rm\cite{Kantor3.5} (Section 4, Prop.~12.)}
\label{sats2}
\!\!\!\!\!Assume that $g$ is {finite-dimensional} and simple. Then there 
are nonzero elements $e,\,f,\,h$, at level one, minus one and zero, respectively, that satisfy the {\rm Chevalley relations}
\begin{align}
[h,\,e]&=2e\,, & [h,\,f]&=-2f\,, & [e,\,f]&=h\,.
\end{align}
\end{thm}
Thus 
$e,\,f,\,h$
are the Chevalley basis elements
corresponding to a simple root.
Theorem 1 and 2 together give the following corollary.
\newtheorem{cor}[thm]{Corollary} 
\begin{cor}
If $T$ is $K$-simple and antisymmetric in the first and third arguments, then $g$ is infinite-dimensional.
\end{cor}
\def\Pf{\noindent \textbf{Proof. }}
\Pf
Suppose the contrary, that $g$ is finite-dimensional. It follows by Theorem \ref{sats1} that
$g$ is simple. But then according to Theorem \ref{sats2} there are nonzero elements $e,\,f,\,h$ such that
\begin{align} 
[[e,\,f],\,e]=[h,\,e]=2e\,.
\end{align}
On the other hand we have
\begin{align} 
[[e,\,f],\,e]=(e\tau(f)e)=0
\end{align}
since the triple product is antisymmetric in its first and third arguments.
Thus we get a contradiction and we conclude that $g$ is infinite-dimensional.
\def\qed{\hspace{\stretch{1}} $\square$ \\
\noindent}
\qed

In particular, for the totally antisymmetric triple system used in the BLG theory with eight supersymmetries, the associated Lie algebra $g$ is infinite-dimensional. Indeed, this triple system is K-simple since the structure constants are proportional to the $so(4)$ epsilon tensor.
We stress that although $g$ is infinite-dimensional, each of the infinitely many subspaces $g_k$ is finite-dimensional. Again, $g$ should not be confused with its subalgebra $g_0$ (which is $so(4)$ in this case), nor with the triple system itself (which can be identified with $g_1$). 
 
There is still the possibility that $g$ is an infinite-dimensional Kac-Moody algebra. From Theorem 2 we only know that in the finite-dimensional case, it is possible to find a Chevalley basis and a simple root such that the corresponding elements $e,\,f,\,h$ belong to level one, minus one, and zero, respectively. If $g$ is an infinite-dimensional Kac-Moody algebra then
the grading cannot be given by a simple root in this way. It might be possible to find elements $e,\,f,\,h$ at level one, minus one, and zero, respectively, such that $e$ and $f$ are eigenvectors to the adjoint action of $h=[e,\,f]$. But then the eigenvalues must be zero instead of $\pm 2$. This suggest that $g$ is Borcherds algebra \cite{borcherds}, or some even more general algebra that (unlike a Kac-Moody algebra) allows for such zero eigenvalues.




We are finally able to express the ABJM action completely in
terms of the associated graded Lie algebra $g$. We recall that
$Z^A,\,\Psi_A$ are elements in $g_1$, while $Z_A,\,\Psi^A$ are
elements in $g_{1}$, which are mapped onto $Z^A,\,\Psi_A$ under the
involution,
\begin{align}
Z^A &= Z^A{}_aT^a, & Z_A &= Z_A{}^aT_a,\nn\\
\Psi_A &= \Psi_{Aa}T^a, & \Psi^A &= \Psi^{Aa}T_a,
\end{align}
and $A_{\mu}$ belongs to the $g_0$ subalgebra:
\begin{align}
A_{\mu} = A_{\mu}{}^a{}_b S^b{}_a.
\end{align}
The Lagrangian (\ref{lagrange}) can thus be rewritten as
\begin{eqnarray}
{\cal L} &=& -\kappa( D^\mu \bar{Z}_A,\, D_\mu Z^{A}) - i \kappa(
\bar \Psi^{A} ,\, \gamma^\mu
 D_\mu  \Psi_{A} )
 \nonumber\\
 &&  + i \kappa( [\bar {\Psi}^{A},\,\bar{Z}_{B}],\,
 [\Psi_{A},\,Z^B])
 -2i \kappa( [\bar {\Psi}^{A},\,\bar{Z}_{A}],\,
 [\Psi_{B},\,Z^B])
  \nonumber\\
 &&
 -\tfrac{i}{2}\epsilon_{ABCD} \kappa( [\bar {\Psi}^{A},\,{\Psi}^{B}],\,
 [Z^C,\,Z^D])
 -\tfrac{i}{2}\epsilon^{ABCD} \kappa( [\bar Z_A,\,\bar Z_B],\,
 [\bar {\Psi}_{C},\,{\Psi}_{D}]
 )
 \nonumber\\
&& -V +\epsilon^{\mu\nu\lambda}\big(\kappa(
\partial_{\mu}A_\nu ,\, A_\lambda ) -\tfrac{2}{3} \kappa(
[A_{\mu},\,A_\nu],\,A_\lambda ) \big)\,,
\end{eqnarray}
where, after using (\ref{sexindex}), the potential takes the simple form
\begin{eqnarray}
V&=&\kappa( [[\bar Z_A,\,\bar Z_B],\,
Z^C],\,[[Z^A,\,Z^B],\,\bar Z_C] ) \nn\\&&-\tfrac13\kappa( [[\bar Z_A,\,\bar Z_B],\,\bar Z_C],\,[[Z^A,\,Z^B],\,Z^C] ).
\end{eqnarray}
One natural generalization would be to let $Z^A$ and $\psi_A$ take
values in $g_k$ for all positive levels $k$, instead of just $g_1$. This does
not, however, seem to be compatible with supersymmetry.

\section{Conclusions and comments}\label{concl}

This note is based on the observation that the $\mathcal N=6$ ABJM
theory can be written in terms of four-index structure constants
$f^{ac}{}_{bd}$ which are antisymmetric only in the upper pair and
the lower pair separately. The fundamental identity then takes the
same form as the basic identity in a generalized Jordan triple
system suggesting a connection to graded Lie algebras associated to
such triple systems. To rewrite the theory, we use an involution and
an invariant bilinear form on the Lie algebra, which naturally
induce a metric on the generalized Jordan triple system. However,
this means that we do not need to use the  metric explicitly in
constructing the Lagrangian.

We have been very general in the description of the Lie algebra
associated to a generalized Jordan triple system. The example
that it first of all should be applied to is the three-algebra
given by Bagger and Lambert in \cite{Bagger:2008se}.
The relation between their work and ours should be studied in detail.
Also, the position of the indices suggests an interesting connection to
the embedding tensor method used in \cite{Bergshoeff:2008bh}.

Even if much of what we have presented in this note are based on
reformulations of previous results, we think that our approach opens
up new perspectives. We have interpreted the fields $Z^A,\,\Psi_A$
as elements in $g_1$, their conjugates as elements in $g_{-1}$, and
the gauge field $A_{\mu}$ as an element in $g_0$. Although we do not
have any interpretation of the elements at higher (positive and
negative) levels, we cannot set them to zero, because we need the
triple product to be antisymmetric in the first and third argument.
Therefore we believe that also the full algebra might play an
important role in the theory of M2-branes. For example, it points
out a new direction in which one could possibly search for the
behavior $n^{3/2}$ that the degrees of freedom of $n$ M2-branes are
conjectured to exhibit. In any case, it would be interesting to see
how fast the dimension of the Lie algebra grows as we go to higher
levels. The algorithm that we have described for finding the
corresponding $g_0$-representations would probably be easy to
implement in a computer program.

There are many implications following from a relation between
M2-brane systems and generalized Jordan triple systems. In
particular, very little is known about the structure of such triple
systems when the grading is infinite. Finite-dimensional cases are
better known and many of their properties have been studied (for an
overview of Jordan, Kantor and Freudenthal triple systems, we refer
to \cite{Palmkvist:2005gc}). For instance, in analogy with
Freudenthal triple systems (see e.g. \cite{Gunaydin:2000xr}), we may
suspect that the generalized Jordan triple systems used here might also be of interest
in connection with minimal representations, spherical vectors and
the associated automorphic forms. For previous attempts to use the
theory of automorphic forms in the context of the M2-brane, see
\cite{Pioline:2001jn,Pioline:2004xq}.

Let us end by mentioning two other issues. The Freudenthal triple
system construction leads to minimal representations via non-linear
realizations of the full algebra \cite{Gunaydin:2000xr}. In
\cite{Aharony:2008ug} the authors argue that the M2-theory discussed
here really has eight supersymmetries but that the last two are
somehow realized non-locally. The connection to triple systems may
in fact suggest how to derive non-linear realizations also of the
remaining two supersymmetries needed to obtain the maximal number of
$\mathcal N=8$ supersymmetries.

The second issue is the one of unitarity. Standard triple system
constructions naturally lead to Lie algebras that appear in their
split form although other forms are also possible. To achieve
unitarity one may try to quantize the theory whereby an
infinite-dimensional unitary minimal representation is realized on a
Hilbert space. For an explicit example, see \cite{Gunaydin:2001bt}.


\acknowledgments

We would like to thank Joakim Arnlind, Ling Bao, Ulf Gran, Andreas Gustavsson, Carlo Meneghelli, Christoffer Petersson and Hidehiko Shimada for discussions. The work is
partly funded by the Swedish Research Council.


\end{document}